\documentclass[showpacs,superscriptaddress,twocolumn,amsmath,aps]{revtex4}
\usepackage{graphicx,color}
\usepackage{amsfonts,amssymb}
\usepackage{bm}
\usepackage[hypertex]{hyperref}

\newcommand{\be}{\begin{equation}}
\newcommand{\ee}{\end{equation}}
\newcommand{\bea}{\begin{eqnarray}}
\newcommand{\eea}{\end{eqnarray}}
\newcommand{\bsube}{\begin{subequations}}
\newcommand{\esube}{\end{subequations}}

\newcommand{\Eq}[1]{Eq.\,(\ref{#1})}

\newcommand{\dg}{\dagger}
\newcommand{\la}{\langle}
\newcommand{\ra}{\rangle}

\newcommand{\ti}{\Tilde}
\newcommand{\nl}{\nonumber \\}

\newcommand{\beq}{\begin{equation}}
\newcommand{\eeq}{\end{equation}}
\newcommand{\beqn}{\begin{eqnarray}}
\newcommand{\eeqn}{\end{eqnarray}}
\newcommand{\bsub}{\begin{subequations}}
\newcommand{\esub}{\end{subequations}}

\newcommand{\ket}[1]{{\left| #1 \right\rangle }}

\begin{document}
%\begin{CJK*}{GBK}{song}
%--------------------------------------------------------------

\title{Nonadiabatic molecular dynamics simulation: An approach
based on quantum measurement picture}

\author{Wei Feng}
\affiliation{Department of Physics, Beijing Normal University,
Beijing 100875, China}
\author{Luting Xu}
\affiliation{Department of Physics, Beijing Normal University,
Beijing 100875, China}
\author{Xin-Qi Li}
\email{lixinqi@bnu.edu.cn}
\affiliation{Department of Physics, Beijing Normal University,
Beijing 100875, China}
\affiliation{Key Laboratory of Theoretical and
Computational Photochemistry of Ministry of Education,
Beijing Normal University, Beijing 100875, China}

\author{Weihai Fang}
\affiliation{Key Laboratory of Theoretical and
Computational Photochemistry of Ministry of Education,
Beijing Normal University, Beijing 100875, China}
\affiliation{Department of Chemistry, Beijing Normal University,
Beijing 100875, China}
\author{YiJing Yan}
\affiliation{Department of Chemistry, Hong Kong University
      of Science and Technology, Kowloon, Hong Kong}

\date{\today}

%% \maketitle
\begin{abstract}
Mixed-quantum-classical molecular dynamics simulation implies
an effective quantum measurement on the electronic states
by the classical motion of atoms.
Based on this insight, we propose a quantum trajectory mean-field
approach for {\it nonadiabatic} molecular dynamics simulations.
The new protocol provides a natural interface
between the separate quantum and classical treatments,
without invoking artificial surface hopping algorithm.
Moreover, it also
bridges two widely adopted nonadiabatic dynamics methods,
the Ehrenfest mean-field theory and the trajectory surface-hopping method.
Excellent agreement with the exact results is illustrated
with representative model systems, including the challenging ones
for traditional methods.
\end{abstract}

\pacs{03.65.Yz,03.65.Sq,31.15.xv,31.15.xg}

\maketitle

\section{Introduction}

A full quantum mechanical evaluation for molecular dynamics (MD)
would quickly become intractable with the increase of atomic degrees of freedom.
As alternatives in practice,
some mixed-quantum-classical (MQC) MD approaches
were developed and proved to be very powerful \cite{Tul98a,Bar11}.
A typical class of such studies is nonadiabatic MD.
Nonadiabatic effects are of crucial importance
in the proximity of conical intersection,
see Fig.\ 1(A), where the energy separation between different
potential energy surfaces (PESs) becomes comparable with
the nonadiabatic coupling. %%==
MQC treatment of nonadiabatic MD has a long history.
The widely applied schemes include the so-called  ``Ehrenfest'' or
``time-dependent-Hartree mean-field'' approach
\cite{Rat82,Mic83,Fri05},
the ``trajectory surface-hopping'' methods
\cite{Tul71,Mi72,Kun79,Bla83,Mi83,Tul90},
and their mixed scheme \cite{Kun91,Ros94,Ros97,Zhu04}.
The former views that the electronic wave function
is in general a linear combination of Born-Oppenheimer adiabatic states,
and the atomic effective potential (and force) is calculated
by averaging the electronic Hamiltonian over such wave function.
The trajectory surface-hopping scheme is in a different extreme.
It believes that the trajectories should split into branches,
i.e., each trajectory should be on one state or another,
not somewhere in between.
In this type of theories the trajectories distribution is achieved
by allowing hops between PESs according to some probability distribution.

\begin{figure} %[!htbp]
%\centering
\includegraphics[width=6cm]{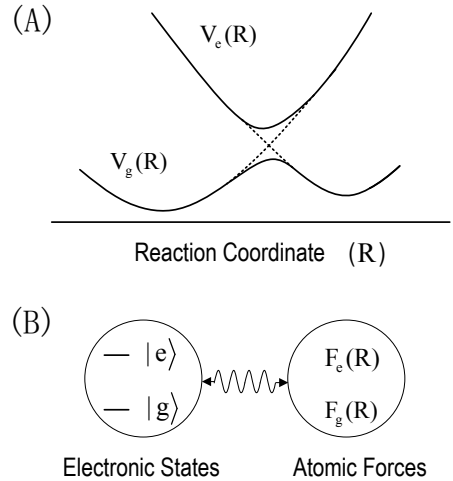}
  \caption{ (A) Schematic atomic potentials in terms of
adiabatic (solid) and diabatic (dashed) representations
for the electronic ground and first excited states
along the reaction coordinate.
(B) Measurement analogy by regarding the atomic motion as
quantum measurement which continuously probes the electronic
states by, for instance, the distinct forces experienced. }
\end{figure}

Among the trajectory-based surface hopping methods, the most popular
one is Tully's
{\it fewest switches surface hopping} (FSSH) approach \cite{Tul90},
together with its variations \cite{Bar11}.
In this approach, nonadiabatic dynamics is treated
by allowing hops from one PES to another,
with the hopping probability determined by
the weight change of the respective electronic states
that are in a {\it quantum superposition}.
From observation that the {\it classical}
atomic motion must decohere the electronic subsystem
(from quantum superposition), considerable efforts were pushed
towards accounting for the associated decoherence effect
\cite{Sch96,Tha98,Pre99,Sch05,Zhu05,Gra07,Sub11}.

One may notice that the FSSH treatment has an obvious flaw
from basic physical point of view.
By an analogy with quantum measurement or
the popular Schr\"odinger's Cat paradox, as illustrated in Fig.\ 1(B),
the FSSH scheme simply indicates that, while having found the Cat
definitely ``alive'' or ``dead'', one is still
treating the radioactive decay in ``quantum superposition''.
In addition, it was also noticed that
the FSSH scheme involves a man-made hopping
algorithm to generate the stochastic surface-switching events
\cite{Bar11,Pre99,Sch05}.
In this work, by explicitly identifying the role of the atomic motion
as a quantum measurement to the electronic subsystem,
we propose a novel quantum trajectory
mean field (QTMF) approach to nonadiabatic MD simulations.
The protocol is developed from an insight that the involved
quantum weak measurement actually
serves as an interface between the {\it quantum}
and {\it classical} parts of the MQC strategy.
Our scheme also naturally unifies the Ehrenfest-type
mean field theory and the trajectory surface-hopping method.
As illustrated by excellent agreement with
the exact results on the three representative models
discussed by Tully \cite{Tul90}, the QTMF approach can
eliminate all the unsatisfactory features of the FSSH method.

%\vspace{0.2cm}
%%%=================
\section{Formulation}

 Let us start with the electronic Hamiltonian
%(including the nuclear potential),
\be\label{Hel}
H_{el}(r;\textbf{R}) = -\sum_j \frac{\hbar^2}{2m_j}\nabla_j^2
   + v(r;\textbf{R}) .
\ee
The potential $v(r;\textbf{R})$, which includes the nuclear potential,
depends on electronic coordinates operator $r\equiv \{r_j\}$
and also atomic configuration, $\textbf{R}\equiv \{R_\alpha(t)\}$,
that is assumed a set of classical dynamics variables.
Expand the electronic wavefunction with an orthogonal basis set functions,
$\Psi(r,t;\textbf{R})=\sum_jc_j(t)\phi_j(r;\textbf{R})$.
In particular one often exploits
the Born-Oppenheimer (BO) adiabatic wavefunctions.
%$\{\phi^{\text{\tiny BO}}_j(r;\textbf{R})\}$.
These are the instantaneous
eigenstates of $H_{el}(r;\textbf{R})$,
satisfying $H_{el}(r;\textbf{R})\phi_j(r;\textbf{R})
= \varepsilon_j(\textbf{R})\phi_j(r;\textbf{R})$,
the standard output of quantum chemistry computation.
Each BO energy serves as
the BO potential energy surface (PES),
$\varepsilon_j(\textbf{R})\equiv {V}_j(\textbf{R})$, for nuclear motion.

Without loss of generality, we proceed
with the BO representation hereafter.
The Schr\"odinger equation for the
{\it coherent} electronic evolution reads  %%==
\be\label{Seq-1}
 \dot{c}_j = -\frac{i}{\hbar} V_j(\textbf{R})\, c_j
  - \sum_k \dot{\textbf{R}}\cdot\textbf{d}_{jk}(\textbf{R})\,c_k ,
\ee
with the nonadiabatic coupling characterized by
%%%
\be \label{eq2}
\textbf{d}_{jk}(\textbf{R}) = \la \phi_j(r,\textbf{R})|
\nabla_{\textbf{R}} |\phi_k(r,\textbf{R})\ra \, .
\ee
%%%
Treating atomic motion with classical trajectories
on individual PESs, $\{V_j(\textbf{R})\}$,
the highly celebrated FSSH method \cite{Tul90}
goes by a Monte-Carlo algorithm as follows.  %%==
It uses \Eq{Seq-1} for the hopping probability from a given
${V}_j(\textbf{R})$ to another.
That is $p_j(t)=\left[|c_j(t)|^2-|c_j(t+\Delta t)|^2\right]/|c_j(t)|^2$,
the normalized population change in the
BO electronic state $|\phi_j(\textbf{R})\ra$.
%%%
However, this algorithm is of {\it problematic} basis.
It completely neglects the influence
of classical trajectories back onto
the electronic state evolution.

Atomic motion that experiences a series of single PESs should
\emph{collapse} the electronic state from a quantum superposition,
given by \Eq{Seq-1}, onto the corresponding single BO basis state,
due to the entanglement-type correlation between the two subsystems.
%%==
In other words, atomic motion is continuously extracting
{\it information} on the electronic state,
via the correlation between the PES and BO basis state.
For instance, the force experienced by atomic motion
plays essentially the same role
as the meter's {\it output} in quantum measurement process. 
Based on this insight, we propose to apply
the well-established quantum trajectory equation,
in replacement of \Eq{Seq-1},
to account for the {\it backaction effect} of the atomic ``meter''
on the electronic subsystem \cite{WM10}:
\begin{align}\label{QTE}
\dot{\rho}(t)&=-\frac{i}{\hbar}  \left[{H}_{el}(\textbf{R}),\rho(t)\right]
  +\sum_{j\neq k} \Gamma_{jk}
 {\cal D}\left[M_{jk}(\textbf{R})\right]\rho(t) \nl
&\quad + \sum_{j\neq k} \sqrt{\gamma_{{\rm F},jk}+\gamma'_{jk}}
{\cal H}\left[M_{jk}(\textbf{R})\right]\rho(t)\xi_{jk}(t) .
\end{align}
In this equation, $\rho$ denotes the reduced density matrix
of electronic state, with diagonal elements for BO-state
population probabilities, and off-diagonal ones for their coherence.
The first term in \Eq{QTE} describes
the same coherent dynamics of \Eq{Seq-1},
corresponding to the Ehrenfest mean-field approach.
The second term accounts for the decoherence effect
owing to ensemble average of the nuclear degrees of freedom,
with an overall rate $\Gamma$.
The third term, significantly, reflects the {\it backaction effect}
of the atomic motion in each single trajectory realization,
with a rate $\gamma_{\rm F}$
for force-mediated information gain
and $\gamma'$ for information gain by other means,
e.g., the nuclear coordinate (atomic configuration)
and velocity (atomic kinetic energy).
Here we omitted the PES indices of the rates
for brevity and for a general description.
In the following Sec.\ II (A) and (B) we will explain
how these rates can be implemented in MQC-MD simulation.
Before that, we describe in more detail the decoherence
and measurement backaction terms.

The second decoherence term in \Eq{QTE} is associated
with a Lindblad superoperator
${\cal D}[M_{jk}]\rho = M_{jk}\rho M_{jk}^{\dg}
 - \frac{1}{2}\{ M_{jk}^{\dg}M_{jk}, \rho\}$,
where $M_{jk}(\textbf{R})=|\phi_j(\textbf{R})\ra  \la \phi_j(\textbf{R})|
- |\phi_k(\textbf{R})\ra  \la \phi_k(\textbf{R})|$
indicates a dephasing between states
$\phi_j(\textbf{R})$ and $\phi_k(\textbf{R})$.
The last backaction term, explicitly,
is described in terms of an superoperator as
${\cal H}[M_{jk}]\rho = M_{jk}\rho + \rho M_{jk}^{\dg}
- \la M_{jk}+M_{jk}^{\dg} \ra \rho $, where
$\la M_{jk}+M_{jk}^{\dg}\ra\equiv {\rm Tr}[(M_{jk}+M_{jk}^{\dg})\rho]$.
Involved in \Eq{QTE} for this back-action effect
are also the quantum jump (from the Copenhagen postulate)
related stochastic noises, $\{\xi_{jk}(t)\}$,
which satisfy the ensemble average property of
${\rm E}[\xi_{jk}(t)\xi_{j'k'}(t')]=\delta_{jk,j'k'}\delta(t-t')$.
From the quantum trajectory theory \cite{WM10},
the last term in \Eq{QTE} has a role of collapsing
the electronic state from a quantum superposition
onto a single BO basis state.
Therefore, now, the issue on ``devising" hopping algorithms
that are not contained in \Eq{Seq-1} does no longer exist anymore.

\subsection{Information Gain Rates}

In the MQC-MD approach, the nuclear part is treated classically.
As a consequence, just like Tully pointed out
in his pioneering work \cite{Tul90},
the classical {\it force} experienced by the atomic motion
in the {\it no-transition adiabatic area} should come from a single PES.
This indicates that, from a measurement perspective,
the classical force plays a role of measurement {\it output}. 
Below we analyze this {\it force-mediated}
information gain rate ($\gamma_{\rm F}$).

We know that the emergence of {\it classicality} from
a closed quantum system is a fundamental puzzle in quantum mechanics.
In essence, this transition is accompanied by {\it quantum jumps}.
This implies that the classical force has certain stochastic fluctuations.
%%%
Since the atomic motion is much slower than its electronic counterpart,
it would be reasonable to use a {\it coarse-graining} force
to evolve the Newton equation.
Let us denote the coarse-grained fluctuating component by $\ti F_j(t)$,
which is an average over a characteristic time ``$\tau_c$''
around $t$ as follows:
\bea
\tilde{F}_j(t)=\frac{1}{\sqrt{\tau_c}}
\int^{t+\tau_c/2}_{t-\tau_c/2}
{\rm d}t'\xi_j(t')\bar{F}_j .
\eea
$\bar{F}_j$ is the BO force associated with the $j_{\rm th}$ PES
at $\textbf{R}(t)$.
Notice also that $\xi_j{\rm d}t={\rm d}W_j$, the Wiener increment,
has a magnitude order and dimension of $\sqrt{{\rm d}t}$ \cite{WM10}.
As a result, the coarse-grained $\tilde{F}_j(t)$ is
no longer $\delta$-function correlated, but
has a correlation function of
\begin{equation}
{\rm E}[\tilde{F}_j(t)\tilde{F}_j(0)]
= \left\{
  \begin{array}{cc}
    (\bar{F}_j)^2 (\tau_c-t)/\tau_c, &  ~~ 0<t<\tau_c \\
    (\bar{F}_j)^2 (\tau_c+t)/\tau_c, &  ~ -\tau_c<t<0
  \end{array}
\right.
\end{equation}
where ${\rm E}[\cdots]$ denotes an ensemble average
over the stochastic realization $\xi_j$, which satisfies
${\rm E}[\xi_j(t)\xi_j(t')]=\delta(t-t')$.
Accordingly, the zero-frequency spectrum of the
force-force correlation function reads
\bea
S_j=\int^{\infty}_{-\infty} {\rm d}t ~{\rm E}
[\tilde{F}_j(t)\tilde{F}_j(0)] = (\bar{F}_j)^2 \tau_c.
\eea

Now we return to the original (stochastic) force of the
$j_{\rm th}$ PES, ${\cal F}_j(t)=\bar{F}_j(t)+\sqrt{S_j}\xi_j(t)$.
The total average force $F_j$, given by averaging ${\cal F}_j$ 
over time interval $(t,t+\tau_m)$, is a stochastic variable
satisfying a Gaussian distribution
$P(F_j)=(2\pi D_j)^{-1/2}\exp[-(F_j-\bar{F}_j)^2/2D_j ]$,
with the variance given by $D_j=S_j/\tau_m$.
Following the theory for realistic
quantum measurements \cite{Mtime},
the state distinguishable criterion
\bea\label{mtime}
\sqrt{D_j}+\sqrt{D_k} \leq |\bar{F}_j-\bar{F}_k|  ,
\eea
allows us to extract the measurement time ($t_m$)
which is minimally required
to identify the state being $\ket{\phi_j}$ or $\ket{\phi_k}$.
Obviously, $t_m$ is given by $\tau_m$ from \Eq{mtime}
in the case of equality.
Then, the information gain rate $\gamma_{{\rm F},jk}$ in \Eq{QTE}
coincides with $1/t_m$, taking a compact form as,
\be\label{Gam_in_F}
\gamma_{{\rm F},jk}=\frac{(\bar{F}_j-\bar{F}_k)^2}
{(|\bar{F}_j|+|\bar{F}_k|)^2} \frac{1}{\tau_c}\, .
\ee
As inferred from the coarse-grained force,
the characteristic time $\tau_c$ physically
scales atomic motion that is typically of picosecond.
With respect to the femtosecond timescale of the electronic part,
in practice we adopt $1/\tau_c\simeq 10^{-3}$ in (reduced) units
of the electronic energies.
Favorably, the information gain rate given by \Eq{Gam_in_F}
is of configuration ($\textbf{R}$) dependence,
but it does not need any microscopic
information of the nuclear (quantum) wavepackets.
It allows thus for a convenient implementation
even for simulation on complex molecular systems.

Except for the force-mediated information gain discussed above,
there exit also other channels of information gain,
which are formally accounted for by $\gamma'$ in \Eq{QTE}.
The channels include the nuclear coordinate $\textbf{R}$
and velocity $\dot{\textbf{R}}$ in the MQC-MD simulation.
For instance, if we performed a microscopic full quantum treatment,
the nuclear wavepacket would have distinct spatial
extension along different PES.
With this ``knowledge" in priori, one can infer certain information
for the electronic state from the classical ``output" $\textbf{R}$.
Another information channel is the nuclear kinetic
energy (associated with $\dot{\textbf{R}}$).
For an initial state with specific energy,
the distinct kinetic energies on different PESs
in MQC-MD simulation exposure also
some information of the electronic state.
Unfortunately, to our knowledge, the information gain rates
through these channels have not yet well developed so far.
However, fortunately, as we will elaborate further
in next subsection, the specific form of these rates
are not important for us to get the correct results.

\subsection{More Remarks on Eqs.\ (\ref{QTE}) and (\ref{Gam_in_F})}

%At this position, it might be useful to make
%some complementary remarks on \Eq{QTE}.
%%
Generally speaking, a desirable MQC-MD approach
should satisfy two requirements. One is that the equation
for the electronic subsystem should satisfy
the ensemble average property, corresponding to averaging
the nuclear degrees of freedom from the exact dynamics
of the full electronic-plus-nuclear system.
Another is that the equation should allow for performing
reasonable classical MD simulation
(with correct ``force" and ``kinetic energy")
on the nuclear subsystem.
Our protocol is to combine \Eq{QTE} with a classical MD simulation
to fulfill these two requirements.
That is, the second term of \Eq{QTE} satisfies
the first requirement, and the last term satisfies the second one.

In \Eq{QTE}, we distinguished the {\it information-gain}
rates $\gamma_{\rm F}$ and $\gamma'$
from the {\it overall} decoherence rate $\Gamma$.
Formally, we may express
$\Gamma=\gamma_{\rm F}+\gamma'+\tilde{\gamma}$.
As discussed above in Sec.\ II (A), $\gamma_{\rm F}$
and $\gamma'$ are, respectively, the information
gain rates mediated by force and other channels
(e.g., $\textbf{R}$ and $\dot{\textbf{R}}$).
Therefore, $\tilde{\gamma}$ represents the decoherence rate
{\it not} unraveled (the rate of information loss).
%%%%
For practical applications,
we propose the following strategies to implement these rates:

\begin{enumerate}

\item
In the nonadiabatic crossing region,
use the rate $\gamma_{\rm F}$ given by \Eq{Gam_in_F} to
approximate the total information gain and decoherence rates.
This approximation is from an insight that, in the conical
intersection area, the total decoherence rate should
be quite weak, otherwise the result will be strongly
distorted from the correct one
(the Ehrenfest mean-field approach is a support to this viewpoint).
We believe that our coarse-graining argument for obtaining
\Eq{Gam_in_F} gives a reasonable order of magnitude for this rate.

\item
Apart from the crossing area, add a nonzero $\gamma'$
into \Eq{QTE}, by using a simple phenomenological parameter
with similar/stronger magnitude order of $1/\tau_c$,
or by certain more sophisticated manner \cite{Sch05,Sub11,Nitz93}.
We may remark that this different implementation
of $\gamma'$ is anticipated to result in slight difference
{\it only} in the ``narrow" region between
the nonadiabatic crossing and the no-transition adiabatic areas.
It will affect neither the molecular dynamics in the broad
adiabatic region, nor the ensemble statistical properties.
For the overall decoherence rate $\Gamma$,
one can set either $\Gamma=\gamma_{\rm F}+\gamma'$,
or $\Gamma=\gamma_{\rm F}+\gamma'+\tilde{\gamma}$
by including a more information-loss rate.
However, $\tilde{\gamma}$ will have no effect,
since $\gamma_{\rm F}$ and $\gamma'$ will collapse
the system {\it in the adiabatic area}
onto a single PES in each trajectory realization,
implying a mixed state after ensemble average.
The role of $\tilde{\gamma}$ is simply to facilitate
the formation of an ensemble-averaged mixed state.

\end{enumerate}

Finally, we mention a special case that may remind our attention.
For parallel PESs, the ``force output" reveals no information
of the electronic state, thus giving a vanished measurement rate.
This is in consistence with the result of \Eq{Gam_in_F}.
In this case, the rate $\gamma'$ from other informational channels
will collapse the system onto a single PES {\it in the adiabatic area}.
Whether or not collapsing the system onto a single PES in this case
will have no effect on the force, but it {\it does} affect
the kinetic energy (nuclear velocity) that should be
of importance in the MD simulation for real molecular systems.

\subsection{Issue of Energy Conservation}

In the MQC-MD approach, the atomic motion defines
a time-dependent electronic Hamiltonian,
which does not conserve the electronic energy.
In turn, the electronic energy defines a potential
to guide the classical motion of atoms.
The sum of the kinetic and potential energies,
$E = K + {\rm Tr}[H_{el}(\textbf{R})\rho(\textbf{R})]
\equiv K + V(\textbf{R})$, is conserved,
simply as the situation in the Ehrenfest mean-field approach.

However, \Eq{QTE} is stochastic. This would lead
to a stochastic potential energy $V(\textbf{R})$.
The non-analytic $V(\textbf{R})$ makes the force
not perfectly defined in mathematics, causing thus
some errors in determining the nuclear velocity.
This would violate slightly the total energy conservation.
Noticeably, in the present QTMF approach, this violation is quite weak
(particularly if a coarse graining procedure is involved), unlike the
drastic violation in the FSSH scheme where an {\it energy calibration}
must be performed after each hopping event.
For practice of the QTMF approach,
we propose the following scheme to address this issue:
\begin{enumerate}
    \item Define the whole simulation region
with the criterion $V(\textbf{R})\leq E_0$,
where $E_0$ is the initial energy of the whole system.
Of course, this renders also that $V_j(\textbf{R})\leq E_0$
once the system is fully collapsed onto the $j_{\rm th}$-PES.
    \item If $V(\textbf{R}_1)= E_0$ occurs at $\textbf{R}_1$,
reset the system to its proximity point $\textbf{R}_2$,
given by $V_M(\textbf{R}_2)= E_0$.
Here, $V_M$ is the renormalized Ehrenfest mean-field potential
energy and determined as follows:
at $\textbf{R}_2$, keep the electronic wavefunction
unchanged as the one at $\textbf{R}_1$;
subtract the lowest PES component
and re-normalize the wavefunction;
then use the renormalized wavefunction to calculate
the Ehrenfest $V_M(\textbf{R}_2)$.
    \item Restart the MD evolution from
the determined proximity point $\textbf{R}_2$,
with the original superposition of BO PESs at $\textbf{R}_1$
%with the unchanged electronic wavefunction at $\textbf{R}_1$,
but a newly assigned atomic kinetic energy of $E_0-{V}(\textbf{R}_2)$
and the momentum direction opposite to that at $\textbf{R}_1$.
\item
After passing through the nonadiabatic crossing area, check the total energy
of the collapsed state (onto a single PES) and make it be $E_0$
via proper modification to the kinetic energy.
\end{enumerate}

%Fig. 2:  =========================================
\begin{figure} %[!htbp]
%  \centering
  \includegraphics[width=7.5cm]{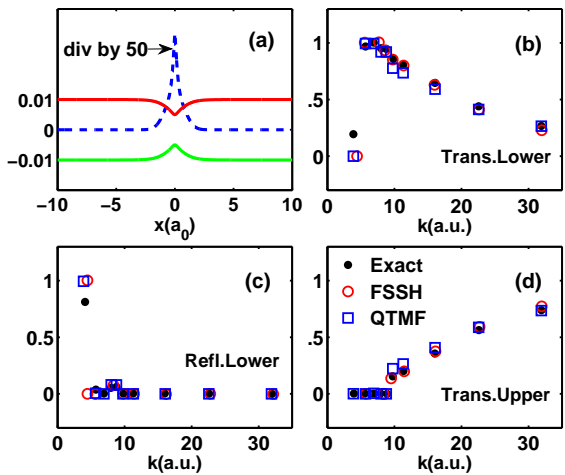}
  \caption{ Single-avoided crossing model.
  In (a) we depict the adiabatic potential sufaces (solid)
  and the nonadiabatic coupling strength (dashed,
  given by \Eq{eq2}).
  Displayed in (b), (c) and (d) are, respectively, the probabilities
  of transmission to the upper surface, reflection to the lower surface,
  and transmission to the upper surface.
  For comparison, we put together the results of our QTMF and Tully's
  FSSH approaches against with the exact one of quantum dynamics simulation.  }
\end{figure}

%% ==================
%\vspace{0.1cm}

\section{Illustrative Examples}

In this section we present our QTMF results
{\it versus} the exact and FSSH counterparts,
on the well-known set of Tully test systems \cite{Tul90},
each of them being a one-dimensional two-surface model,
with an atomic mass of $M=2000$\,a.u.\
(all parameters below are in atomic unit).
The scheme for exact quantum dynamics simulation
was clearly described in Ref.\ \cite{Tul90}.
In the present work, we simply extract the results
from Ref.\ \cite{Tul90} for comparison.
In our simulation, we assume an incident energy
$E_0=k^2/(2M)$ to initiate the system evolution.
And, as discussed earlier, we adopt $1/\tau_c=10^{-3}$.
In the nonadiabatic coupling area, we approximate
the entire decoherence and information rates
by $\gamma_{\rm F}$ through \Eq{Gam_in_F}.
In the adiabatic (no-transition) area, we add $\gamma'=10^{-2}$
to account for the backaction effect of other informational channels,
and set $\Gamma=\gamma_{\rm F}+\gamma'$.
As explained in Sec.\ II (A) and (B), the choice
of $\gamma'$ in the adiabatic area can be rather arbitrary,
having almost no influence on the results.
For each model, we run $2000$ stochastic trajectories and accordingly
determine the population probabilities of the final ``products''.
Also, each trajectory begins with the classical particle
(atom) on the lower energy surface at $x=-10$, with an incident
momentum to the right, and ends at $|x|=15$.

Model-I: {\it Single-Avoided Crossing} --
The diabatic potential matrix elements for this model are
\be\label{tully-1}
\begin{split}
 &V_{11}(x) = -V_{22}(x) = \text{sign}(x)\cdot A [1-\exp(-Bx)],
\\
 &V_{12}(x) = V_{21}(x) = C \exp(-Dx^2).
\end{split}
\ee
%%%
Set $A=0.01$, $B=1.6$, $C=0.005$, and $D=1$.
The adiabatic potential surfaces of this model are plotted in Fig.\ 2(a),
while the results are shown in Fig.\ 2(b)-(d).
%%
%For the purpose of comparison, here and in the following
%(the other two models), we put together also the results
%from the FSSH approach and the wavepacket-evolution-based
%{\it exact} quantum dynamics simulation.
Desirably, both our QTMF and the FSSH schemes work very well for
this model, being almost in an overall agreement with the exact results.
We only make two remarks on the extremely quantum regime.
{\it (i)}
The steep step-function behavior at $k\sim 5$ is an indicator
for the failure of almost all {\it semiclassical} MD methods,
i.e., failing to predict tunneling through the barrier
on the lower surface at very low momentum.
Physically, in our case this is caused by setting the {\it semiclassical
rule} of energy-conservation when propagating the state. Giving up this
restriction at $k\sim 5$, we can actually recover the exact result.
{\it (ii)}
Another interesting quantum regime is $k\sim 8$ ($7.7<k<8.9$),
which is above the threshold for transmission.
Satisfactorily, both QTMF and FSSH captured the essential physics here,
e.g., predicting the small amount of particle reflections.
This is somehow a challenging test for any semiclassical approaches.

%Fig. 3:  =========================================
\begin{figure} %[!htbp]
%  \centering
  \includegraphics[width=7.5cm]{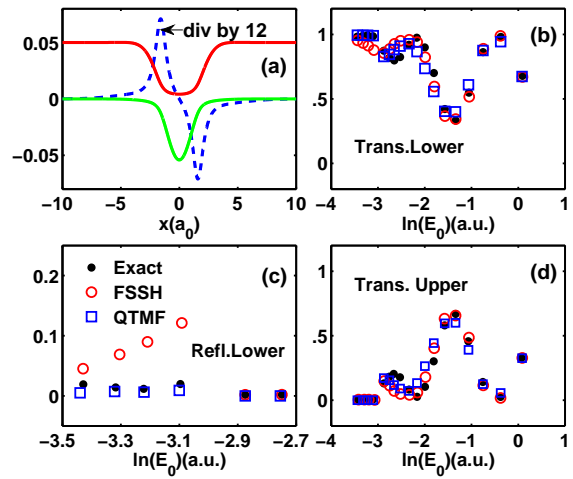}
  \caption{
  Dual-avoided crossing model. Shown in (a) is the
  adiabatic potential surfaces (solid) together with
  the nonadiabatic coupling strength (dashed),
  while in (b), (c) and (d)
  are the transmission and reflection probabilities
  as stated in Fig.\ 2.   }
\end{figure}

%Fig. 4:  =========================================
\begin{figure}  %%%[!htbp]
%  \centering
  \includegraphics[width=7.5cm]{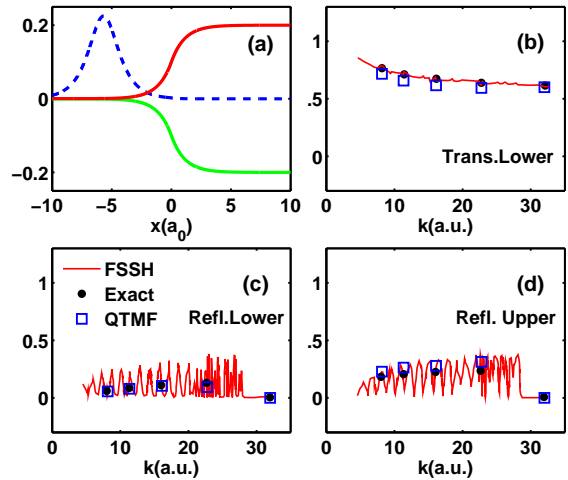}
  \caption{  Extended coupling model.
  Shown in (a) is the
  adiabatic potential surfaces (solid) together with
  the nonadiabatic coupling strength (dashed),
  while in (b), (c) and (d)
  are the transmission and reflection probabilities
  as stated in Fig.\ 2.  }
\end{figure}
%%%

%\vspace{0.2cm}
Model-II: {\it Dual-Avoided Crossing} --
This is a more challenging model and contains two avoided crossings.
The key feature of
this model is the Stueckelberg oscillations, owing to quantum
interference between the successive nonadiabatic quantum transitions.
The diabatic potentials for this model are given by
\be\label{tully-2}
\begin{split}
 &V_{11}(x) = 0, \ \ \  V_{22}(x) = -A\exp(-Bx^2)+E , \\
 &V_{12}(x) = V_{21}(x) = C\exp(-Dx^2),
\end{split}
\ee
where $A=0.10$, $B=0.28$, $C=0.015$, $D=0.06$, and $E=0.05$.
The adiabatic potentials of this model
and results comparison are shown in Fig.\ 3.
At high incident energies, both the FSSH and QTMF can give
correct results in good agreement with the exact ones,
all exhibiting the expected Stueckelberg oscillations.
%%\{{\bf high incident energy = larger value $\ln E$ (or $\ln E_0$?)}\}
However, at low energies, the FSSH method fails to predict both
the transmission and reflection probabilities on the lower surface,
see Fig.\ 3 (b) and (d) in the low energy regime.
In particular, the FSSH algorithm overestimates the amount of reflection
by an order of magnitude (roughly a factor of 10).
This overestimation is owing to the {\it artificial} hopping algorithm,
which results in too many upward transitions. In sharp contrast,
our QTMF approach can physically rule out this drawback.

%\vspace{0.2cm}

Model-III: {\it Extended Coupling} --
This is the most challenging model for classical mechanics
based approach to address, since it involves an extended region
of strong nonadiabatic coupling. The diabatic potentials are
\be\label{tully-3}
\begin{split}
  &V_{11} = - V_{22} = A  ,
\\
  &V_{12} = V_{21} = \begin{cases} B\exp(C x);&  x<0,
\\ B[2-\exp(-C x)];&   x>0. \end{cases}
\end{split}
\ee
The parameters are $A=6\times 10^{-4}$, $B=0.1$ and $C=0.9$.
The adiabatic potentials and comparative results are shown in Fig.\ 3.
We see that, unfortunately, the FSSH algorithm
completely fails for the reflection probabilities,
to either the upper or lower surface.
This failure clearly indicates that the FSSH algorithm
breaks down in strong quantum transition region.
Again, in sharp contrast, our QTMF approach gives
satisfactory results even for this very demanding model.

\section{Summary}

%\vspace{0.2cm}
%%  ========================================================
To summarize, we have proposed a quantum trajectory mean field (QTMF) approach
to the powerful mixed-quantum-classical molecular dynamics simulation
with surface hopping.
Simulations on three nontrivial models are quantitatively satisfactory.
While \Eq{Gam_in_F} offers a compact position-dependent
measurement rate on atomic motion timescale ($\tau_c$),
the present study reveals also an important observation:
results are rather insensitive to the choice of decoherence rate,
as long as it is weak ($\sim 1/\tau_c$) in the nonadiabatic crossing area.
Unraveling any decoherence rate in the no-transition adiabatic area
can stochastically collapse the system
onto a single potential surface,
and gives about the same satisfactory statistical results.

In this context we would like to remark that
quantum superposition is {\it rooted}
in the exact quantum dynamics simulation,
but involving not at all the concept of classical atomic trajectory.
%%%
In Tully's fewest switches algorithm, while the evolution of electronic state
is not consistent with the successive {\it complete} surface hopping,
it keeps the essential feature of quantum superposition.
It is merely this reason, in our opinion, that makes
the most celebrated FSSH approach be often comparable
to the exact results from full quantum dynamics simulation.

Compared with the FSSH approach, the QTMF scheme
is founded on a more physical and simpler treatment.
For the electronic part, the replacement of the Schr\"odinger equation
with a master equation will increase only negligible amount
of computational complexity, since the involved BO states are very few 
(for instance, only two in most real molecular simulations).
On the other hand, the QTMF scheme avoids the hopping algorithm
and simplifies the procedures of calibrating the total energy.
This will speed the simulation.
As a conservative estimate,
the time cost of the QTMF scheme is comparable to
or less than the FSSH approach (and its many variations).
With this computational efficiency
together with the satisfactory accuracy,
and most importantly its physical foundation,
the proposed QTMF scheme is anticipated
to be a powerful tool in real MD simulations.

%\vspace{0.5cm}
%% =========================================================
%% \acknowledgments
{\it Acknowledgments}---
This work was supported by the Major State Basic Research
Project of China (Nos.\ 2011CB808502 \& 2012CB932704)
and the NNSF of China (Nos.\ 101202101 \& 10874176 \& 21033008).

%---------------²Î¿¼ÎÄÏ×-----------------------
%\newpage

%\end{CJK*}
\end{document}